\documentclass[aps,prb,preprint,preprintnumber,showpacs]{revtex4-1}
\usepackage{amsmath}
\usepackage{amssymb}
\usepackage{graphics}
\usepackage{epsfig}
\usepackage{color}

\begin{document}



\title{Superconductivity and Charge Density Wave in ZrTe$_{3-x}$Se$_{x}$}
\author{Xiangde Zhu$^{1,2}$, Wei Ning$^{1}$, Lijun Li$^{2}$, Langsheng Ling$^{1}$, Ranran Zhang$^{1}$, Jinglei Zhang$^1$, Kefeng Wang$^{2,\S}$, Yu Liu$^{3}$, Li Pi$^{1,4}$, Yongchang Ma$^{2,5}$, Haifeng Du$^{1}$, Minglian Tian$^{1,4}$, Yuping Sun$^{1,3,4}$, Cedomir Petrovic$^{2}$, Yuheng Zhang$^{1,4}$}
\affiliation{$^{1}$High Magnetic Field Laboratory, Chinese Academy of Sciences and University of Science and Technology of China, Hefei 230031, China\\
$^{2}$Condensed Matter Physics and Materials Science Department, Brookhaven
National Laboratory, Upton, New York 11973, USA\\
$^{3}$Key Laboratory of Materials Physics, Institute of Solid State Physics Chinese Academy of Sciences, Hefei 230031, China\\
$^{4}$Collaborative Innovation Center of Advanced Microstructures, Nanjing 210093, China\\
$^{5}$School of Materials Science and Engineering, Tianjin University of Technology, Tianjin 300384, China}

\date{\today}

\begin{abstract}
Charge density wave (CDW), the periodic modulation of the electronic charge density, will open a gap on the Fermi surface that commonly leads to decreased or vanishing conductivity. On the other hand superconductivity, a commonly believed competing order, features a Fermi surface gap that results in infinite conductivity. Here we report that superconductivity emerges upon Se doping in CDW conductor ZrTe$_{3}$ when the long range CDW order is gradually suppressed. Superconducting critical temperature $T_c(x)$ in ZrTe$_{3-x}$Se$_x$ (${0\leq}x\leq0.1$) increases up to 4 K plateau for $0.04$$\leq$$x$$\leq$$0.07$. Further increase in Se content results in diminishing $T_{c}$ and filametary superconductivity. The CDW modes from Raman spectra are observed in $x$ = 0.04 and 0.1 crystals, where signature of ZrTe$_{3}$ CDW order in resistivity vanishes. The electronic-scattering for high $T_{c}$ crystals is dominated by local CDW fluctuations at high temperures, the resistivity is linear up to highest measured $T=300K$ and contributes to substantial in-plane anisotropy.
\\
\\
Correspondence and requests for materials should be addressed to Y.H.Z (zhangyh@hmfl.ac.cn) and C.P. (petrovic@bnl.gov).
\end{abstract}

\maketitle

Charge density wave (CDW) and superconductivity (SC), both Fermi surface instabilities and low-temperature collective orders in solids, are commonly believed to compete with each other \cite{Gruner,Gabovich}. Recently, dynamic CDW fluctuations have also been discussed in copper oxide superconductors \cite{NM} in connection with quantum critical transition between CDW and superconductivity. CDW favors low dimensional systems, especially transition metal (M) MX$_2$ and MX$_3$ chalcogenides (X represents S, Se and Te) \cite{Gruner,Wilson}. Among them, ZrTe$_3$ is of interest since its crystal structure (Fig. 1a) is quasi-two dimensional (2D), yet it contains two quasi-one dimensional (1D) trigonal prismatic ZrTe$_6$ chains with inversion symmetry along the $b$-axis \cite{Felser}. From the view along $c$-axis (Fig. 1b), the top Te2/Te3 atoms form a rectangular network with the distances of 0.279/0.310 nm along the $a$ axis and 0.393 nm along the $b$-axis. The first principle calculation gives evidence that the electron-type band (Te2/Te3 5$p_x$ in origin) provides the major contribution, whereas the contribution of the partially filled hole-type band that originates in Te1 5$p_y$ and Zr 4$d_{y^2}$ orbitals is minor at the Fermi surface \cite{Felser}. Angular resolved photoemission (ARPES) demonstrates that CDW originates from the Te2/Te3 5$p_x$ band \cite{Hoesch}.

ZrTe$_{3}$ features not only a CDW transition temperature ($T_{CDW}$)$~\sim$ 63 K with a CDW vector $\vec{q}~\approx$  ($\frac{1}{14}$; 0; $\frac{1}{3}$ ) but also a nearly isotropic in-plane and quasi-two-dimensional (2D) electronic transport \cite{Felser,Eaglesham,Takahashi}. There is a filamentary SC in a stoichiometric single crystal with higher onset of $T_{c}$ for $a$-axis from resistivity measurement than for $b$-axis \cite{Felser,Yamaya}. Heat capacity data suggest that SC transitions in ZrTe$_{3}$ are successive from filamentary-to-bulk with local pair fluctuations above T$_{c}$; SC phase first condenses into filaments along a-axis, becoming phase coherent below 2 K\cite{Yamaya}. Pressure($P$), intercalation, and disorder can tune ZrTe$_3$ into bulk SC with suppression of CDW order \cite{Yomo,Zhu,Zhu2}.

Here we provide evidence for the pronounced upper critical field $H_{c2}(T)$ anisotropy and emerging 1D electronic transport along the ZrTe$_6$ chain-direction $b$ axis in ZrTe$_{3-x}$Se$_x$ (${0\leq}x\leq0.1$). The $H_{c2}(T)$ anisotropy and new Raman modes suggest coexistence of local CDW modes and enhanced superconducting $T_c(x)$ in ZrTe$_{3-x}$Se$_x$.

\section*{Results}

Normalized $\rho_a$ [$\rho_a$/$\rho_a$(300K)] [Fig. 1(c)] shows that the CDW anomaly is suppressed with increasing Se content. The $T_{CDW}$ [Fig. 1(d)] decreases whereas the bulk superconductivity sets in [Fig. 1(e,f)]. For $x \leq 0.04$, as shown in Fig. 1(e), the superconducting transition temperature ($T_c$) determined from the $\rho_a (T)$ curves tends to increase, and transition width decreases. With increasing Se content $x \geq 0.04$, [Fig. 1(f)], somewhat lower $T_{c}$ = 4 K is observed for $x$ = 0.07, and the superconducting transition becomes wide and filamentary for $x$ = 0.1 with $T_{c}~\sim$ 3.1 K. ZrTe$_{2.96}$Se$_{0.04}$ shows typical behaviour of a type-II superconductor, whose field cooling (FC) $\chi$ is smaller than in zero field cooling process (ZFC) [Fig. 1(f)].

The $\rho_a$ and $\rho_b$ are almost identical to each other in ZrTe$_3$ (x = 0), however with increasing Se content $x$, room temperature $\rho_a$ tends to increase, while $\rho_b$ tends to decrease [Fig. 2(a)]. This indicates that ZrTe$_{3-x}$Se$_x$ becomes highly conducting along $b$-axis in the normal state. If ZrTe$_{3-x}$Se$_x$ is an anisotropic superconductor with dominant quasi-1D (super)conductivity along the $b$-axis, upper critical field along $b$ axis ($H_{c2} \parallel b$) should be larger than $H_{c2} \parallel a$, according to the single band anisotropic Ginzburg-Landau theory since

\begin{equation}
\Gamma_{ij} = m^*_{ii}/m^*_{jj} = H_{c2}{\parallel}j/H_{c2}{\parallel}i \sim \sqrt{\rho_i/\rho_j}
\end{equation}

To confirm this, we choose the ZrTe$_{2.96}$Se$_{0.04}$ crystal where the ratio of $\rho_{a}$(T)/$\rho_{b}$(T) is about 10 at 300 K [Fig. 2(a)].  The magnetic hysteresis ($M-H$) loop for ZrTe$_{2.96}$Se$_{0.04}$ [Fig. 2(b) inset] confirms that it is a typical type-II superconductor with some electromagnetic granularity.  In ZrTe$_{2.96}$Se$_{0.04}$, $H_{C2}(T)\parallel a > H_{C2}(T)\parallel b > H_{C2}(T)\parallel c$ relation can be observed [Fig. 2(b)]. This is in contrast to the $b$-axis quasi-1D conductivity in the normal state suggesting multiband effects and/or additional factors that can contribute to mass tensor anisotropy. The upward curvature of $H_{c2}$-T curves implies that the multiband effects should be considered.

The $H_{C2}(T)$ for the two-band BCS model with orbital pair breaking is \cite{Gurevich}:
\begin{eqnarray}
&& a_0 \left[ \ln t + U(h) \right]\left[ \ln t + U(\eta h) \right] +
a_2 \left[ \ln t + U(\eta h) \right] \nonumber \\
&&\quad \quad \quad \quad \quad \quad \quad \quad \quad \quad
+a_1 \left[ \ln t + U(h) \right] = 0, \\
&& U(x) = \psi(1/2+x) - \psi(1/2),
\end{eqnarray}
where $t=T/T_{c}$, $\psi(x)$ is the digamma function, $\eta=D_2/D_1$,
$D_1$ and $D_2$ are band 1 and band 2 diffusivities, $h=H_{c2}D_{1}/
(2\phi_{0}T)$, and $\phi_{0}=2.07\times10^{-15}$ Wb is the magnetic flux quantum. $a_{0}=2${\scriptsize$\mathcal{W}$}$/\lambda_{0}$, $a_{1}=1+\lambda_{-}/
\lambda_{0}$, and $a_{2}=1-\lambda_{-}/\lambda_{0}$, where
{\scriptsize$\mathcal{W}$}$=\lambda_{11}\lambda_{22}-\lambda_{12}\lambda_{21}$,
$\lambda_{0}=(\lambda_{-}^{2}+4\lambda_{12}\lambda_{21})^{1/2}$, and
$\lambda_{-}=\lambda_{11}-\lambda_{22}$. Interband coupling in two bands is given by $\lambda_{12}$ and $\lambda_{21}$ whereas $\lambda_{11}$ and $\lambda_{22}$ are intraband coupling constants in band 1 and 2, When $D_{1}=D_{2}$, this simplifies to the one-band model orbital pair breaking in the dirty limit \cite{WHH}. Dominant intraband (interband) coupling is obtained for {\scriptsize$\mathcal{W}$} $> 0$ ({\scriptsize$\mathcal{W}$} $< 0$). The fits to the multiband model using $\lambda_{i,j}$ ${i,j}={1,2}$ in Table I are excellent [solid lines in Fig.2(b)]. Overall, the fitting results indicate dominant intraband coupling\cite{Jaroszynski,LeiH2}. Interestingly, the $\eta\approx0.10(4)$ suggest different $D_{1}$ and $D_{2}$, i.e. approximately an order of magnitude different carrier mobilities in the two bands. This difference in the intraband diffusivities could be due to differences in scattering or effective masses \cite{Gurevich,Jaroszynski}.

Figure 3(a) depicts the $T(x)$ phase diagram of $T_{CDW}$ and $T_c$ versus Se content $x$ for ZrTe$_{3-x}$Se$_x$. With increasing Se content $x$, the CDW order detected by $a$-axis resistivity anomaly is suppressed, vanishing around $x$ = 0.03. SC $T_{c}$ gradually increases up to the maximum $T_c$ = 4.4 K around $x$ = 0.04. With further increase in Se the superconducting $T_{c}$ appears to have onset near 4 K, becoming much broader and with smaller shielding factor suggesting percolative nature of SC. Even though the sample with higher Se content cannot be grown at present, it is clear that the SC should decrease to $T_{c}=0$ since ZrSe$_3$ is a band insulator with a band gap of 1 eV \cite{Patel}. In contrast, Hf substitution on Zr site in Zr$_{1-x}$Hf$_x$Te$_3$ does not suppress the CDW order, and no SC is observed. This is different from IrTe$_2$, in which only 5d Ir site substitution can suppress the charge/orbital order and induce SC \cite{Yang}. The Hf doping does not alter the Te2/Te3 bands, which explains why Hf doping cannot suppress the CDW order and induce SC.

In what follows we compare the Raman signal of superconducting crystals to Raman signal of pure ZrTe$_{3}$ with long range CDW order. Figure 3(b) depicts the Raman spectra normalized to 86 cm$^{-1}$ mode of ZrTe$_{3-x}$Se$_x$ measured at 5K and 300K with Z(XX)Z polarization for different Se content. Small Se doping should not change the phonon spectrum at the room temperature and indeed, the 300 K spectra nearly overlap with each other. As expected, the Raman spectrum of ZrTe$_3$ measured at 5K is different from the one measured at 300K. Two new modes appear around 115 cm$^{-1}$ and 152 cm$^{-1}$, which we assign to CDW (CDW mode) \cite{Raman1,Raman2}. Periodic lattice distortions in the CDW state will result in the new phonon modes below $T_{CDW}$ and some (CDW modes) can be observed in the Raman spectra, for example in 1T-TiSe$_2$ \cite{Snow} and 2H-NbSe$_2$ \cite{Measson}. The 108 cm$^{-1}$, 140 cm$^{-1}$ and 145 cm$^{-1}$ modes are suppressed to low intensity with small temperature dependent shift at low temperature. The intensity of the two CDW modes 115 at cm$^{-1}$ and 152 cm$^{-1}$ becomes weaker for $x$ = 0.04 and 0.1. It should be noted that CDW modes are detected outside the phase boundary of CDW order. The normalized amplitudes of the two CDW modes exist for $x$ = 0.04 and 0.1, in crystals with no CDW signature in resistivity. This suggests a coexistence of superconductivity and CDW-related lattice distortions.

\section*{Discussion}

Fermi surface of ZrTe$_{3}$ contains multiple bands with both flat and dispersive portions as well as substantial hybridization of high mobility chalcogen-derived bands with low mobility metal-derived bands\cite{Felser}. We note that in ZrTe$_{3}$ CDW fluctuations affect the angular resolved photoemission spectral function $A(k,\omega$) at temperatures above 200 K \cite{Yokoya}. Scattering in such multiband CDW electronic system in the presence of local CDW fluctuations is dominated by scattering off collective CDW excitations below $T_{CDW}$ [$\rho(T)$$\sim$$AT^{2}$; where $A$ is a constant parameter] and the electron-phonon and impurity-like scattering off local CDW fluctuations above the $T_{CDW}$ [$\rho(T)$$\sim$$aT+b$]\cite{Naito}. Both $\rho_{a}$ and $\rho_{b}$ are perfecly linear from about 60 K up to highest measured temperature of 300 K, whereas a $\sim$$T^{2}$ resistivity is observed from superconducting $T_{c}$ up to about 80 K [Fig. 4(a,b)]. With suppression of the CDW order by pressure or doping, CDW mode should vanish\cite{Barath}, and indeed the absence of characteristic CDW-related hump is observed in resistivity. However as evident in Fig. 4(c), the signature of CDW mode in ZrTe$_{3-x}$Se$_x$ appears below about 100 K suggesting that the crystallographic vibration of the unit cell still senses CDW presence, but we speculate with no phase coherence.

The resistivity shows that electron-electron scattering due to CDW fluctuations dominates over the electron-phonon scattering and provides $\rho_{a}$$\sim$$AT^{2}$ temperature dependence with relatively high values of coefficient $A$. As a result, Kadowaki-Woods scaling $A_{a}$/$\gamma^{2}$ is comparable to Tl$_{2}$Ba$_{2}$CuO$_{6}$, Sr$_{2}$RuO$_{4}$ or Na$_{0.7}$CoO$_{2}$ [Fig. 3(a) insets]\cite{KW,Jacko,Nozieres}. In ZrTe$_3$, with increasing $P$, CDW order is first enhanced and reaches maximum $T_{CDW}$ around 2 GPa \cite{Yomo}, then decreases, vanishing around 5 Gpa, whereas superconducting $T_c$ increases monotonically up to highest pressures. The phase diagram in Fig. 4b is different from this but Se could act as a chemical pressure due to its smaller size. Therefore slight increase in band filling of the quasi-1D Fermi surface sheet seen in pure ZrTe$_{3}$ under pressure could be the mechanism for the in-plane anisotropy and promotion of SC \cite{Hoesch2}. However, due to very small Se content (up to about 3 atomic \%) and no appreciable change in the unit cell parameters, this would imply strong sensitivity of CDW to substitutions on Te site and possibly to disorder.

As doping increases beyond $x=0.04$ the superconducing $T_{c}$ forms a weak dome or plateau-like temperature dependence similar to PrFeAsO$_{1-x}$F$_{x}$ \cite{Rotundu}. Charge-mediated attraction is involved in both CDW and SC. For well nested Fermi surface long range CDW is stable and superconductivity is only filamentary along a axis arising in the Te2/Te3 5p$_{x}$ band \cite{Felser,Yamaya}. With Se doping the CDW is no longer detected in scattering but dominant intraband interaction could ensure that patches of CDW still survive, as seen by Raman. The small Se substitution is unlikely to remove the nesting condition but may perturbe the long range phase coherence of CDW and consequently resistivity hump. Broad superconducting transition, small reduction of SC transition temperature and significant decrease in the superconducting volume fraction suggest that the percolative SC is independent of Se content once the CDW-related resistivity anomaly is absent.

The above discussion suggests a possibility for a CDW-fluctuation induced heavy-fermion-like mass enhancement contribution to mass tensor anisotropy \cite{Murray}. Moreover, superconductivity on the verge of the breakdown of the long-range CDW order is reminiscent to magnetic fluctuation mediated superconductivity in copper oxide and heavy fermion materials where the magnetic order is tuned by doping or pressure to $T\rightarrow0$ at the Quantum Critical Point \cite{Monthoux,Si,Sebastian,Cortes}.

\section*{Conclusion}

In summary, we show that superconductivity in ZrTe$_{3-x}$Se$_x$ single crystals arises in the background of CDW fluctuations that contribute to significant anisotropy of the both  normal state resistivity and the upper critical field in the superconducting state. The CDW fluctuations exist outside of the phase boundary of CDW order.

\section*{Methods}
Single crystals of ZrTe$_{3-x}$Se$_x$ were grown via iodine vapor transport method \cite{Zhu}. The as grown single crystals can be easily cleaved along $b$-axis and $c$-axis, which usually produces needle- or tape- like crystals along $b$-axis in the $ab$ plane (shown in inset of Fig. 1c). Elemental analysis was performed by energy-dispersive X-ray spectroscopy (EDS) on an FEI Helios Nanolab 600i to determine the Se content. The Se content in as grown crystal is found to be less than the content in the starting material; measured EDS values are presented in figures. Powder X ray diffraction confirms phase purity however there were no appreciable changes of the lattice parameters (below 0.002 {\AA} for $a,b$ and below 0.005 {\AA} for $c$ lattice parameter), as expected for atomic substitution of up to about 3\%. The crystal size becomes smaller when the Se content $x$ increases, reducing from about 3 x 5 mm$^{2}$ for $x=(0-0.04)$ down to 1.5 x 1.5 mm$^{2}$ in $ab$-plane for $x=0.1$. Magnetization was measured in Quantum Design MPMS-XL-5. Resistance and magneto-resistance were measured by four probe method on Quantum Design PPMS-9 and PPMS-16. Raman spectra were measured on Horiba T64000, with excitation wavelength 647.4 nm and the power density was kept below 20 mW cm$^{-2}$ in order to minimize the heating effects.

\begin{acknowledgments}
We thank Shile Zhang, Zhe Qu, Wenhai Song, Lanpo He, Shiyan Li for help on heat capacity measurements. Work at High magnetic field lab (Hefei) was supported by  National Basic Research Program of China (973 Program), No. 2011CBA00111, and National Natural Science Foundation of China (Grants No. U1432251, 11204312, 11474289). Work at Brookhaven National Laboratory is supported by the US DOE under Contract No. DE-SC00112704.
\end{acknowledgments}

\S{Present address: CNAM, Department of Physics, University of Maryland, College Park, Maryland 20742, USA}

\textbf{Author contributions} \\

X.D. Z., C. P. and Y. H. Z. designed the experiments and wrote the draft. L. P, M. L. T. and Y. P. S. discussed the results and commented on the manuscript. Single crystals growth: X. D. Z., L. J. L. SEM and EDS: H. F. D  Resistivity measurements and fits: W. N., K. F. W., L. J. L., X. D. Z and C. P.. Magnetization: L. S. L., Y. C. M, Y. L. Raman: R. R. Z., X. D. Z. Heat Capacity: J. L. Z

\textbf{Competing financial interests}: The authors declare no competing financial interests.

\newpage

\begin{figure}[tbp]
\includegraphics[scale=0.5]{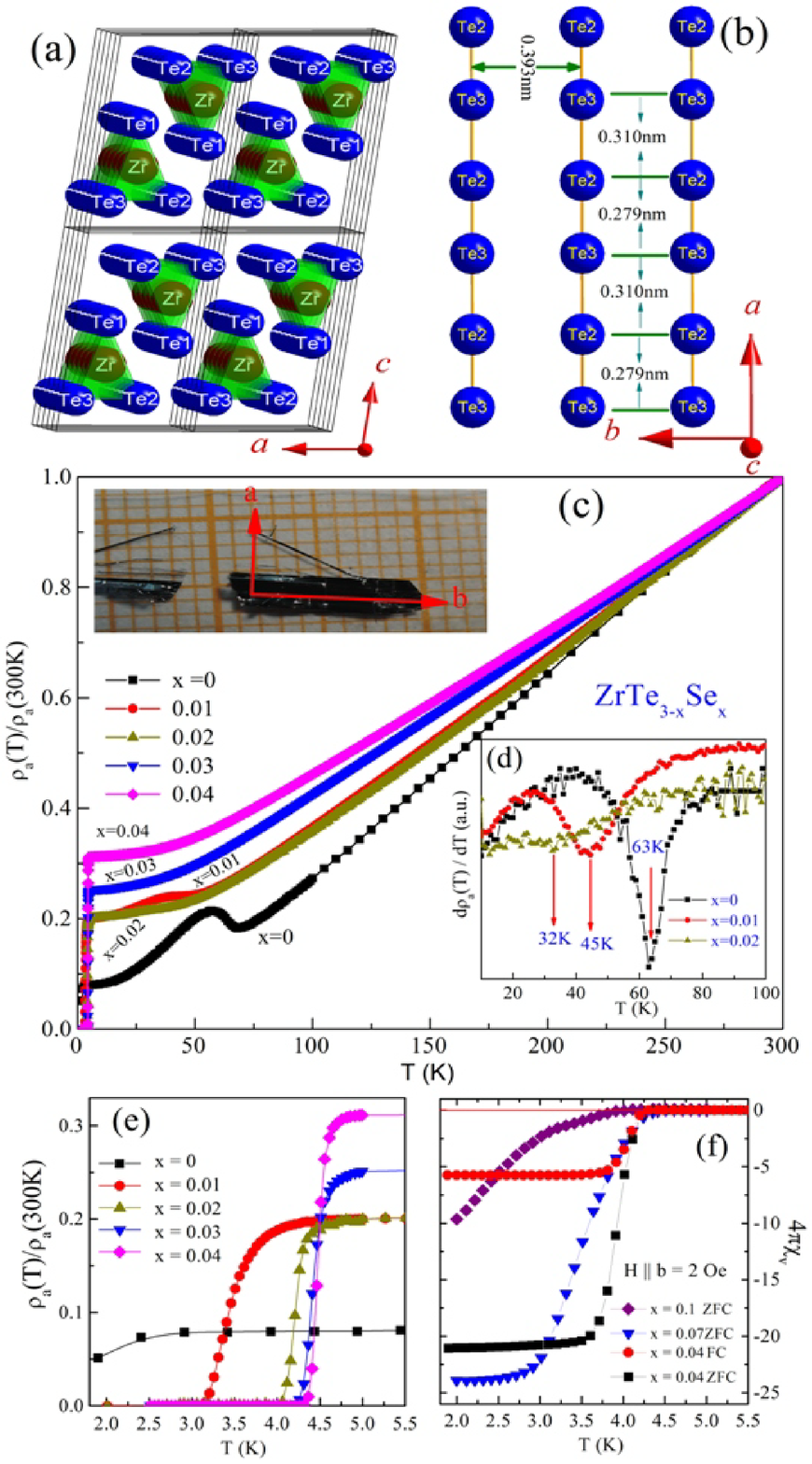}
\caption{(a) Crystal structure of ZrTe$_3$. (b) Top Te2/Te3 rectangular network layer viewed from $c$-axis. The quasi 1D ZrTe$_6$ chains run along the $b$-axis, with the shortest Zr-Zr distance and Te1-Te1 distance of 0.393 nm. Solid line denotes the alternately spaced Te2/Te3 chain. (c) Temperature dependence of normalized $\rho_a$($\rho_a$/$\rho_a$(300K)) for ZrTe$_{3-x}$Se$_x$. The inset shows a typical photograph of cleaved ZrTe$_{3-x}$Se$_x$ crystal. Some fibers along $b$-axis can be observed. (d) The $T_{CDW}$ is determined from the dips in the differential curves of $\rho_a$/$\rho_a$(300K)-$T$ (shown in the inset). Solid rectangular, circle and triangle represent x = 0, 1\% ad 2\%, respectively. The arrows mark the $T_{CDW}$. (e) Low temperature $\rho_a$/$\rho_a$(300K)-$T$. Superconducting $T_c$  is determined as the midpoint of the superconducting transition. \textbf{f}, The temperature dependence of magnetic susceptibility ($\chi$) measured for ZrTe$_{2.96}$Se$_{0.04}$, ZrTe$_{2.93}$Se$_{0.07}$,and ZrTe$_{2.9}$Se$_{0.1}$. The applied magnetic field($H$) is 2 Oe and parallel to the $b$ axis of crystal.}
\end{figure}

\begin{figure}[tbp]
\includegraphics[scale=0.8] {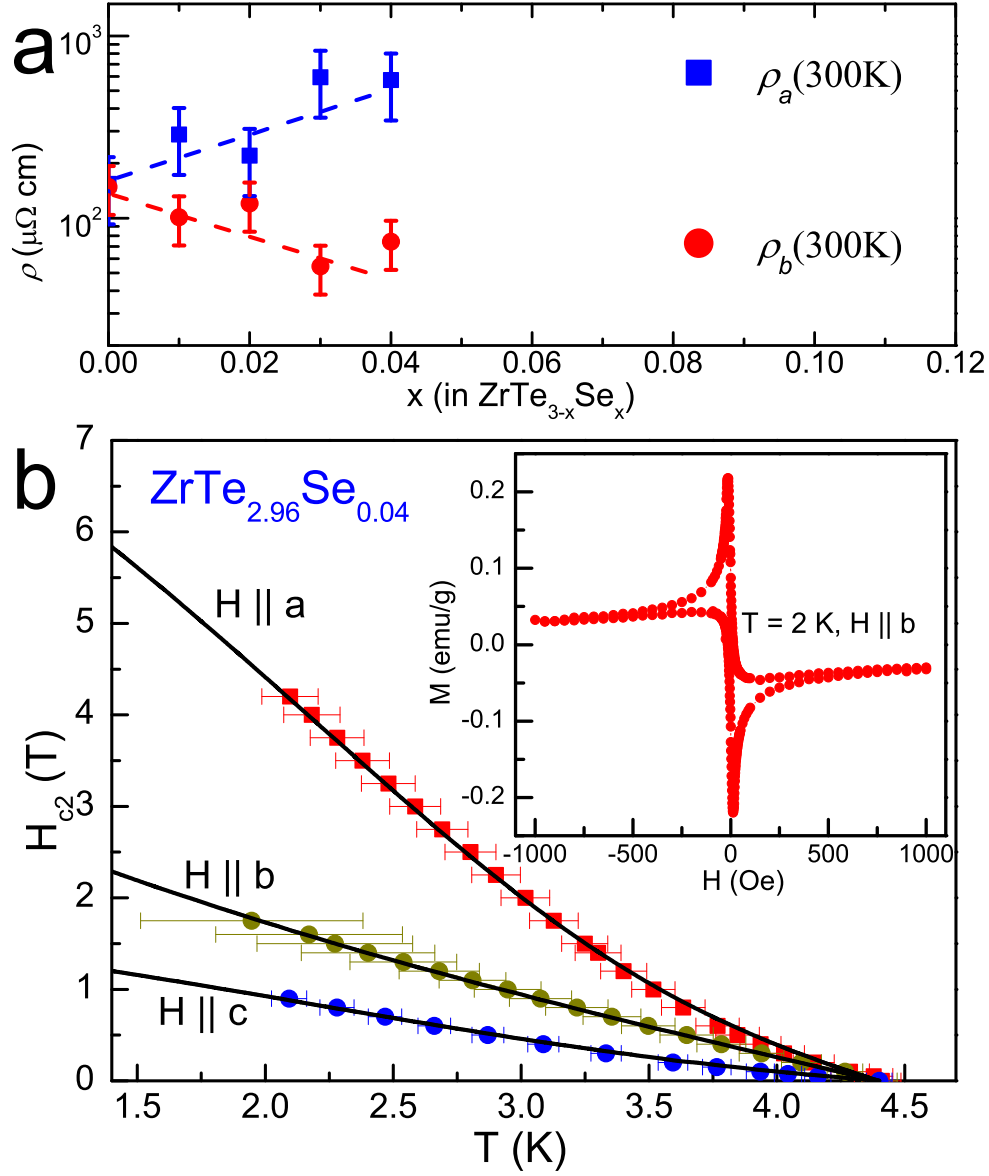}
\caption{(a) The log-plots of $\rho_a$(solid square) and $\rho_b$(solid circle) versus Se doping content. (b) The temperature dependence of the upper critical field (H$_{c2}$) (determined as the midpoint of the superconducting transition; the error bars are the difference from the 10\% and 90\% resistivity drop) of ZrTe$_{2.96}$Se$_{0.04}$ for H $\parallel$a, H $\parallel$ b and H $\parallel$ c. The solid lines represents the fitts of the two band model (see text). Inset shows the magnetic hysteresis ($M-H$) loop for ZrTe$_{2.96}$Se$_{0.04}$ measured at 2K.}
\end{figure}

\begin{figure}[tbp]
\includegraphics[scale=0.7] {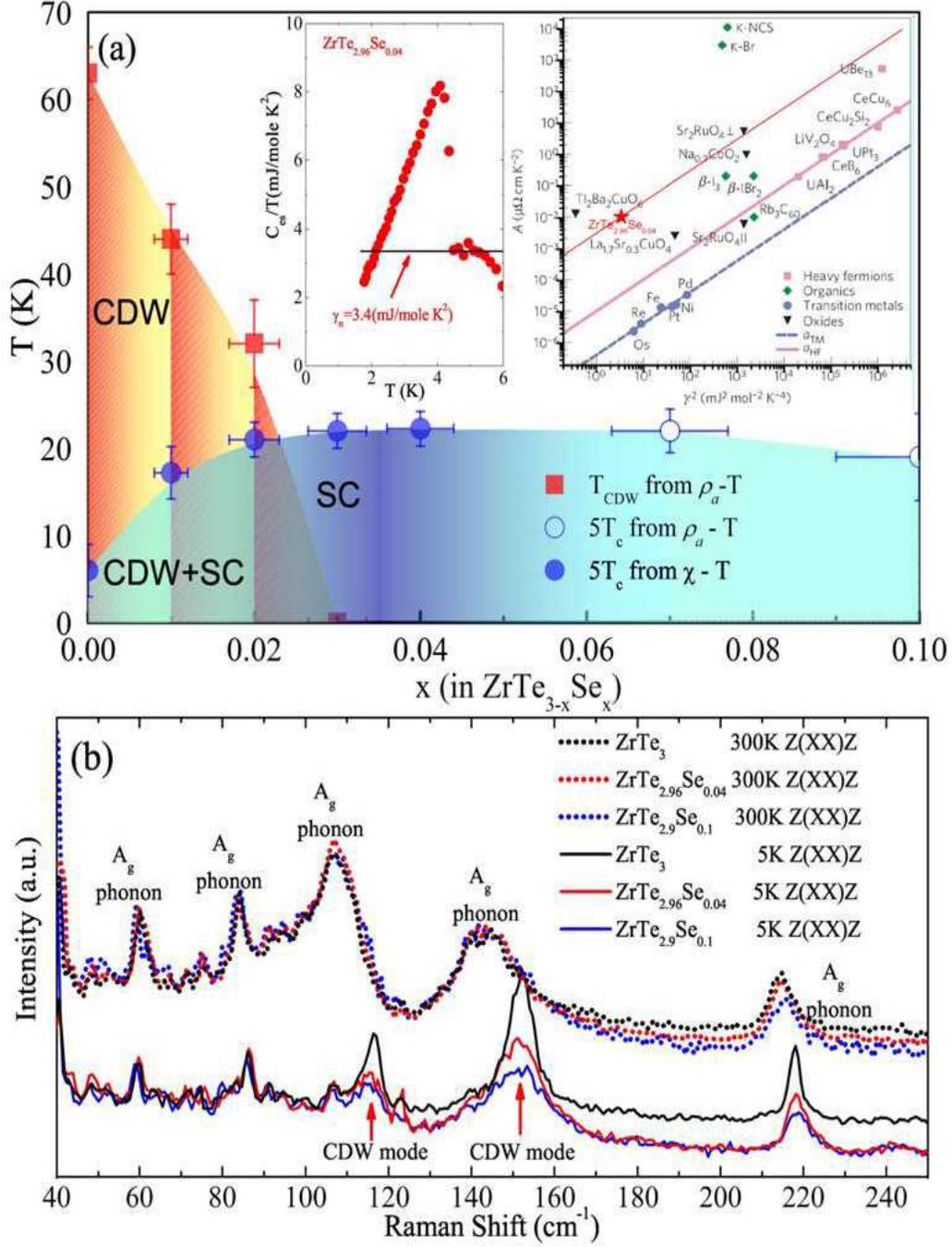}
\caption{(a) The phase diagram of $T_{CDW}$ and $T_c$ versus Se doping content; insets show electronic specific heat and the Kadowaki - Woods ratio for ZrTe$_{2.96}$Se$_{0.04}$ (Refs. 26 and 27). The $a_{TM}$ = 0.4$\mu$$\Omega$cm mol$^{2}$K$^{2}$J$^{-2}$ and $a_{HF}$ = 10$\mu$$\Omega$cm mol$^{2}$K$^{2}$J$^{-2}$ are values seen in the transition metals and heavy fermions, respectively. Even though values of electron-electron scattering rate A and mass renormalization $\gamma$ are smaller than in strongly correlated materials, it appears that the scaling A/$\gamma$$^{2}$ in ZrTe$_{3}$ is similar to Na$_{x}$CoO$_{2}$ and Sr$_{2}$RuO$_{4}$. (b) The normalized Raman scatering spetra for ZrTe$_3$, ZrTe$_{2.96}$Se$_{0.04}$, and ZrTe$_{2.9}$Se$_{0.1}$ measured at 5K and 300K with Z(XX)Z polarization. The two CDW modes at 115 cm$^{-1}$ and 152 cm$^{-1}$ are marked by arrows.}
\end{figure}

\begin{figure}[tbp]
\includegraphics[scale=0.7] {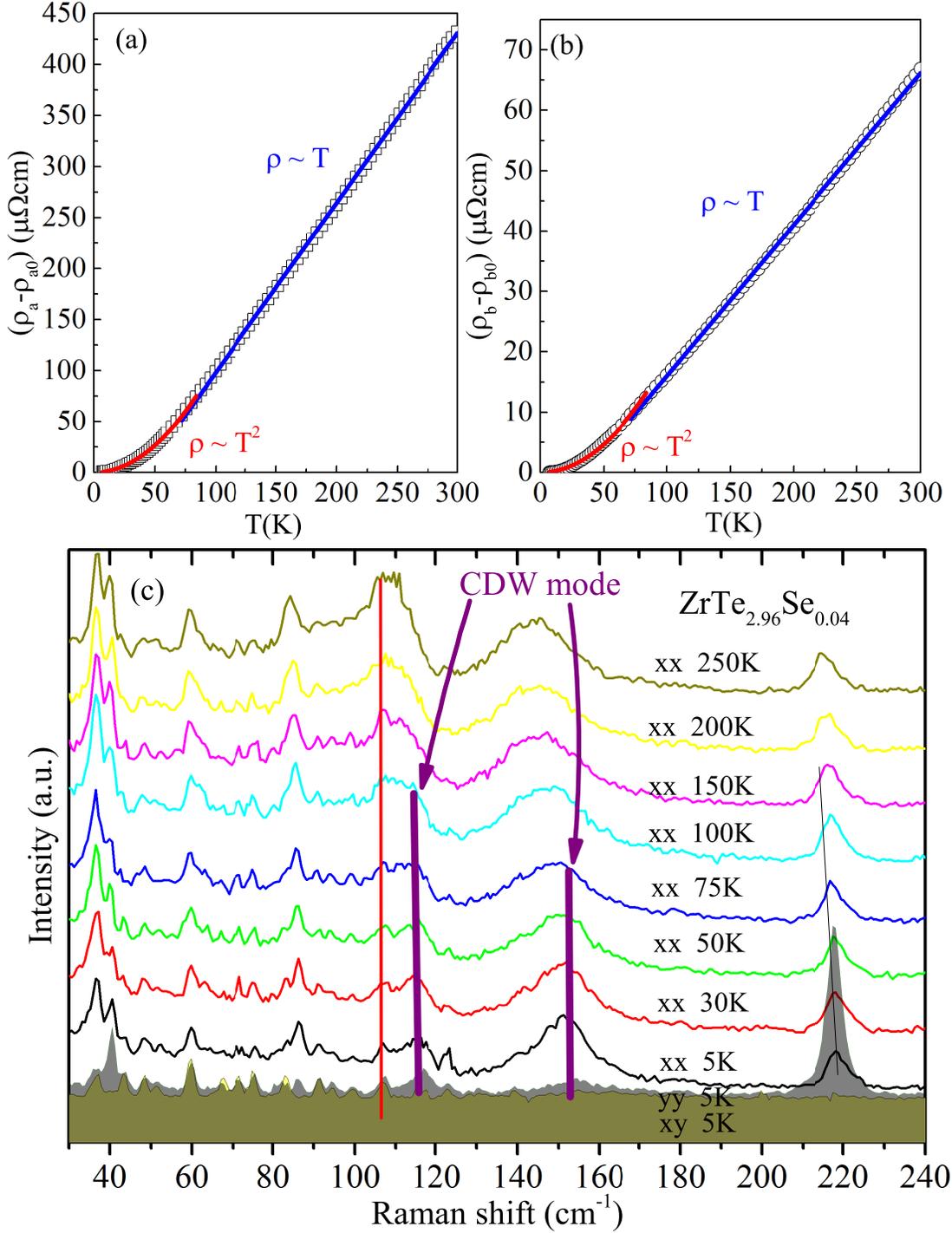}
\caption{(a,b) The a- and b-axis resistivity fits of ZrTe$_{2.96}$Se$_{0.04}$. Below 63 K $\rho(T)$$\sim$$AT^{2}$ and above that temperature $\rho(T)$$\sim$$aT+b$ up to highest measured 300 K. The fitting parameters are $A_{a}$=0.0107(1), $A_{b}$=0.0019(1); $a_{a}$=0.250(1), $a_{b}$=1.66(1) and $b_{a}$=-9.0(2) and $b_{b}$=-68(1). (c) Raman scattering in ZrTe$_{2.96}$Se$_{0.04}$ where CDW mode can be traced below about 100 K.}
\end{figure}

\begin{table}[tbp]\centering%
\caption{Citting parameters of $H_{c2}$ for Cu$_{0.05}$ZrTe$_{3}$.}%
\begin{tabular}{ccccccc}
\hline\hline
$H_{c2}||$ & $\eta$ & $\lambda_{11}$ & $\lambda_{12}$ & $\lambda_{21}$ & $\lambda_{22}$\\
\hline
a & 0.098 & 0.60 & 0.25 & 0.25 & 0.80\\
b & 0.124 & 0.60 & 0.50 & 0.50 & 0.60\\
c & 0.145 & 0.60 & 0.25 & 0.25 & 0.80\\
\hline\hline
\end{tabular}%
\label{CW}%
\end{table}%

\end{document}